\documentstyle[prd,aps]{revtex}
\def\aa{{\cal A}}

\def\l{\label}

\def\th{\Theta}

\def\lan{\langle}
\def\ran{\rangle}
\def\beq{\begin{equation}}
\def\bfg{\begin{figure}}
\def \ts{\textstyle}
\def\efg{\end{figure}}
\def\eeq{\end{equation}}
\def\bea{\begin{eqnarray}}
\def\eea{\end{eqnarray}}

\def\ssec{\subsection}

\begin{document}
\input epsf
\twocolumn[\hsize\textwidth\columnwidth\hsize\csname
@twocolumnfalse\endcsname
\draft
%\rightline{SISSA-xxx/yy/z}
\vskip1pc
\title { {\bf Geometric  reheating after inflation}} 
\author{Bruce A. Bassett \footnote{email:bruce@stardust.sissa.it}
and Stefano Liberati \footnote{email:liberati@sissa.it}}
\address{International School for Advanced Studies, Via Beirut 2-4, 
34014, Trieste, Italy}
\date{\today}
\maketitle
\begin{abstract}
Inflationary  reheating via resonant production of non-minimally coupled 
scalar particles with only gravitational coupling is shown to be extremely 
strong, exhibiting a negative coupling instability for $\xi < 0$ and a 
wide resonance decay for $\xi \gg 1$. Since non-minimal 
fields are generic after  renormalization in curved spacetime, this offers 
a new paradigm in reheating - one which naturally allows for efficient 
production of  the massive  bosons  needed for GUT baryogenesis. 
We  also  show that both vector  and tensor fields are  produced 
resonantly during  reheating, extending the previously known  
correspondences between  bosonic fields of different 
spin during preheating. 
\end{abstract} 
%\vspace*{2cm}
%\centerline{SISSA-ref 120/97/A}
%\vspace*{1cm}
\vskip2pc]
%%%%%%%%%%%%%%%%%%%%%%%%%%%%%%%%%%%%%%%%%%%%%%%%%%%%%%%%%%%%%%%%%%

\section{Introduction}

Typical modern incarnations of inflation arise within supergravity, 
string or GUT theories where the inflaton, $\phi$, is only one of many 
fields.  Studies  of  inflation including couplings to these other fields, 
as required to  reheat the universe after inflation, 
yield extremely  complex dynamics  \cite{CL96} and are little 
investigated  beyond  hybrid models \cite{hybrid}. Here we turn to a 
minimalist view in which preheating \cite{KLS97,pre1,pre2,devega,GKLS97}
occurs with the inflaton coupled only gravitationaly to other fields. We 
call  this {\em geometric reheating} to emphasize its gravitational  
origin. The most powerful  example of this new mechanism is provided by 
non-minimally  coupled fields (section \ref{ssec:nmc}), where the 
strength of the effect is  due to the  congruence of two facts: (1)  the 
Ricci curvature oscillates  during preheating and (2) the  non-minimal 
coupling, $\xi$, is a  free parameter. The first ensures that there is 
resonance, the second that the effect is non-perturbative. 

The possible importance of this effect is motivated by  renormalization group
studies  in curved-spacetime \cite{BOS92,RH97,BDH94}, which have shown that 
even if the bare coupling, $\xi_0$, is minimal, after renormalization $\xi 
\not  = 0$ generically. Since we are particularly interested 
in the preheating realm which occurs when inflation ends near the Planck 
scale, we are near the ultra-violet (UV) fixed points of the 
renormalization group equations.  
While the UV fixed points may correspond to a conformally invariant field
($m=0$, $\xi= {1\over6}$), in different GUT models the 
coupling may also diverge, $|\xi| \rightarrow \infty$, in the UV 
limit \cite{BOS92,PT}. 
In both cases the nature of geometric reheating is very different from the
standard models based on explicit self-interactions or particle-physics
couplings between fields (see e.g. \cite{KLS97}).

Further we shall study the gravitational production of spin 0,1 and
2 particles due to the expansion of  the universe during preheating,
and will show that a unified treatment in terms of parametric resonance
exists.  This is shown by reducing the evolution  equations 
to generalized Mathieu form: 
\beq 
x'' + [A(k) - 2q \cos 2z] x = 0\,,
\l{eq:mathieu}
\eeq
with time-dependent parameters \footnote{While it is known 
that the  Mathieu formulation is insufficient
\cite{devega} in  some respects, and has lead to  the introduction of 
other approximations - principally that of stochastic 
resonance \cite{KLS97} -  the  Mathieu equation remains a powerful
diagnostic test for the  strength of particle production.}.

The Mathieu equation has rapidly growing solutions controlled by the
Floquet index $\mu_k$.
In the case that $1 \gg q > 0$, the Floquet index in the {\em first} 
resonance band of the Mathieu equation is given by  $ \mu_k = 
((q/2)^2 - (2k/m - 1)^2)^{1/2}$ \cite{mac,KLS97}.
This can be extended to give $\mu^N_k$ in the N-th resonance band 
\cite{GMM94} as long as $A > 0$ and $2N^{3/2} \gg q$:
\begin{equation}
\mu^{N}_k = -\frac{1}{2N} \frac{\sin 2\delta}{[2^{N-1} (N-1)]^2} q^N\,,
\label{eq:weaksol}
\end{equation}
where $\delta$ varies in the interval $[-\pi/2, 0]$ and $\mu^N_k \ll 1$.

When $A(k) < 0$ a qualitative change occurs and the dominant effect comes 
from the fact that one effectively has an inverted harmonic oscillator 
yielding the  {\em negative coupling instability} \cite{GPR97}. 
In this case the Floquet index can be as large as $\mu_k \sim 
|q|^{1/2}$, there are no stability 
bands to  speak of and the typical variances are larger by a factor 
$|q|^{1/2}$ than in the  $A(k) > 0$ case.

To be concrete, consider the case of a scalar field in a FLRW universe
($g_{\mu \nu}=diag(-1, a^{2}(t)/(1-Kr^{2}),
a^{2}(t)r^{2}, a^{2}(t)r^{2}sin^{2}\theta)~$, $K = \pm 1,0$ is  the 
curvature constant). We shall
restrict\footnote{We note that application of stochastic
resonance methods to the vector, tensor and non-minimal scalar fields of
this paper for  the potential $V ={\lambda \over 4} \phi^4$ requires an 
extension of the 
existing theory to  scattering in a quartic potential as opposed to the 
standard quadratic potential \cite{KLS97,GKLS97}.} ourselves to the
quadratic potential, 
\beq
V(\phi) = \frac{m_{\phi}^2}{2} \phi^2 \,.
\l{eq:quadpot}
\eeq
For $K=0$, the latter potential gives an oscillatory behaviour of the 
field, $\phi=\Phi \sin(m_{\phi} t)$, with $\Phi \sim 1/m_{\phi}t$.  
In what follows we shall try to preserve  maximal generality; we denote
with $\doteq$ results
which are derived specifically for the potential (\ref{eq:quadpot}). We
use natural units with $\kappa = 8\pi, G = 1$.

The energy density and pressure for a minimally
coupled scalar field, treated as a perfect fluid, are 
$\mu = \kappa({1\over2}\dot{\phi}^2 + V(\phi))$, $p = 
\kappa({1\over2}\dot{\phi}^2 - V(\phi))$.
This breaks down if the field is non-minimally coupled (an imperfect 
fluid treatment must be used), if the effective potential is not adequate 
\cite{devega}, or if large density gradients exist.
The FLRW Ricci tensor is \cite{KS84}
\bea
R^0{}_0 &=& 3\frac{\ddot{a}}{a} \nonumber \\
R^i{}_j &=& \left[\frac{\ddot{a}}{a} + 2\left(\frac{\dot{a}}{a}\right)^2 + 
\frac{2K}{a^2}\right] \delta^{i}{}_j \,,
\l{eq:ricci}
\eea 
where $i,j = 1..3$. The Ricci scalar is:
\beq 
R = 6\left(\frac{\ddot{a}}{a} + \left(\frac{\dot{a}}{a}\right)^2 + 
\frac{K}{a^2}\right) \,.
\l{eq:ricciscalar}
\eeq
The Raychaudhuri equation for the evolution of the expansion 
$\Theta = 3 \dot{a}/a$ is \footnote{The expansion is generally defined as
$\Theta  \equiv u^a{}_{;a}$ where $u^a$ is the 4-velocity and $;$ denotes 
covariant derivative \cite{ellis71}.} given by: 
\beq 
\dot{\th} = -\frac{3\kappa}{2} \dot{\phi}^2+ \frac{3 K}{a^{2}}\,,
\l{eq:ray}
\eeq
while the Friedmann equation is
\beq
\th^2 + {9K\over a^{2}} = 3\kappa \mu = 3\kappa \left({1\over 
2}\dot{\phi}^{2}+  V(\phi)\right)  \,. \l{eq:fried}
\eeq

As an example, when $K = 0$ and $\dot{a}/(a m_{\phi}) \ll 1$, one may 
solve Eq. (\ref{eq:fried}) perturbatively \cite{KH96}, to obtain:
\beq
\Theta \doteq \frac{2}{t}\left[1 - \frac{\sin 2 m_{\phi} t}{2 m_{\phi}
t}\right]\,,
\l{eq:pertth}
\eeq
to first order in $\dot{a}/(a m_{\phi})$. This is only valid after
preheating when $\Phi \ll 1$ but shows that the expansion  oscillates
about the mean Einstein-de Sitter ({\sc  eds}) pressure-free solution. 
Eq.  (\ref{eq:pertth}) can be integrated to give the scale factor:
\beq
a(t) \doteq \overline{a} \exp\left( \frac{\sin 2 m_{\phi}t}{3m_{\phi} t} - 
\frac{2 \mbox{ci}(2 m_{\phi} t)}{3}\right) \l{eq:scale}
\eeq
where $\overline{a} = t^{2/3}$ is the background {\sc eds} evolution, and 
$\mbox{ci}(m_{\phi} t) = -\int^{\infty}_{t} \cos(m_{\phi} z)/z dz$.
This example
explicitly demonstrates how temporal averaging (which yields
$\overline{a}$) removes the resonance. 

Via Eq.'s (\ref{eq:ray},\ref{eq:fried}) one can 
systematically replace all factors of $\dot{a}, \ddot{a}$ with factors of 
$\dot{\phi}$ and $V(\phi)$ 
terms\footnote{Indeed, a useful combination is 
$2\dot{\Theta} + \Theta^2 = -3\kappa p\nonumber 
\doteq \ts{3\over2} \kappa m_{\phi}^2 \Phi^2 \cos (2m_{\phi} t)$
.}. 
In this way one can show that the vector and tensor wave
equations take the form of Mathieu equations during reheating 
\cite{bass97}. 

\section{Scalar fields}

Consider now the effective potential:
\beq
V(\phi,\chi_{\nu})  = V(\phi) + \frac{1}{2}\sum_{\nu}^N m^2_{\nu} 
\chi^2_{\nu} + \frac{1}{2} \sum_{\nu}^N \xi_{\nu} \chi_{\nu}^2 R\,, 
\l{eq:lag} 
\eeq
describing the inflaton with potential $V(\phi)$ coupled only via gravity 
to $N$ scalar fields, $\chi_{\nu}$, which have no self-interactions, masses
$m_{\nu}$ and non-minimal couplings  $\xi_{\nu}$. The equation of motion 
for modes of the $\nu$-th field is: \beq
\ddot{\chi}^{\nu}_k + \Theta \dot{\chi}^{\nu}_k + \left(\frac{k^2}{a^2} + 
m^2_{\nu} +  \xi_{\nu} R\right) \chi^{\nu}_k = 0\,,
\l{eq:nuth}
\eeq
From Eq (\ref{eq:ricciscalar},\ref{eq:ray},\ref{eq:fried}) the Ricci 
scalar  is given by \footnote{Assuming that at the 
start of reheating the inflaton is the dominant contributor to the energy 
density of the universe.}:
\beq
R = - \kappa\dot{\phi}^2 + 4\kappa V(\phi)~~~(K = 0)\,.
\l{eq:ricci2}
\eeq

\ssec{The minimally coupled case}\label{ssec:mcfield}

Consider $\xi_{\nu} = 0$. Then (\ref{eq:nuth}) reduces, on using Eq's  
(\ref{eq:ray},\ref{eq:fried}), to: 
\bea
\frac{d^2 (a^{3/2} \chi_k)}{dt^2} &+& \left(\frac{k^2}{a(t)^2} + 
m_{\nu}^2 + \kappa{3\over8}\dot{\phi}^2 \right.\\ \nonumber
&-& \left. \kappa{3\over4}V(\phi)+{3\over4} {K\over{a^{2}}}\right)(a^{3/2} 
\chi_k) = 0 \,.
\l{eq:vac2} 
\eea
There exists parametric resonance because the expansion $\th$ oscillates. 
The potential (\ref{eq:quadpot}) yields equation (\ref{eq:mathieu}) ($K = 
0$) with time-dependent parameters: 
\beq
A(k,t) \doteq {k^2 \over {a^{2} m^{2}_{\phi} } }+
{m^{2}_{\nu} \over m^{2}_{\phi} }~~~,~~q \doteq {3\over16} \kappa \Phi^2 
\l{eq:mcparam} 
\eeq
~From this we see that the production of particles is reduced as $m_{\nu}$ 
increases. Indeed, since $A \rightarrow m^2_{\nu}/m^2_{\phi}$, $q 
\rightarrow 0$ due to the expansion, production of 
minimally coupled bosons is rather weak and shuts off quickly due to 
horizontal  motion on the stability chart. We stress   that the 
production is, however,  much stronger than that obtained 
in previous studies where the scalar factor evolves monotonically 
\cite{matacz}. This mild situation changes dramatically when a non-minimal 
coupling is introduced.

\ssec{Non-minimal preheating}\label{ssec:nmc}

Now include the {\em arbitrary} non-minimal coupling $\xi_{\nu}$. 
Using Eq. (\ref{eq:ricci2}) one can reduce Eq. (\ref{eq:nuth}) to ($K = 0$):
\bea
\frac{d^2 (a^{3/2} \chi_k)}{dt^2} &+& \left(\frac{k^2}{a(t)^2} +
m_{\nu}^2 + \kappa\left(\frac{3}{8} - \xi\right)\dot{\phi}^2 \right. 
 \nonumber \\ 
&-& \left. \kappa\left(\frac{3}{4}  - 4\xi\right) V(\phi) \right)(a^{3/2} 
\chi_k) = 0 \,. 
\l{eq:nmcvac2}
\eea
Defining a new variable $z=m_{\phi}t+\pi/2$, Eq. (\ref{eq:nmcvac2}) 
takes the form of equation (\ref{eq:mathieu}) with time-dependent 
parameters: 
\bea
A(k,t) &\doteq& {k^2 \over {a^{2} m^{2}_{\phi} } } +
{m^{2}_{\nu} \over m^{2}_{\phi} } + {{\kappa \xi} \over 2}
\Phi^{2}\nonumber\\
q(t) &\doteq& {3\over4} \kappa \left({1\over 4} - \xi \right) \Phi^2 
\l{eq:nmcparam} 
\eea
The crucial observation is that since $\xi$ is initially free to take on any 
value  \footnote{The only constraints that one might impose are that the 
effective potential  should be  bounded from below  and that the strong 
energy condition, $R_{ab} u^a u^b > 0 \Leftrightarrow T_{ab} u^a u^b > 
-T/2$,  be satisfied. The first is difficult to impose since $R$ 
oscillates and the second since one should use the renormalized 
stress-tensor, $\langle T_{ab}\rangle$.}, $A(k)$ is neither 
restricted to be positive nor small.

~From Eq. (\ref{eq:nmcparam}) it is clear that $A(k) < 0$  for 
sufficiently negative $\xi$. The posibility of negative $A$ was the 
thesis of the work by Greene {\em et al} \cite{GPR97}. However, in their 
model, this powerful negative coupling instability was only partially 
effective due to the non-zero vacuum expectation 
value  acquired by the $\chi$ field due to its coupling, $g$, with the 
inflaton. Here we only have gravitational couplings and the same 
constraint is removed. 

Negative $A$ (induced when $q < 0$) implies
that the physical  region of the $(A,q)$ plane is altered. Instead of $A
\ge 2|q|$  we have $A \ge \pi\Phi^2 - 2|q|/3$. Now when $2|q|/3 > |A|
\gg 1$  we have $\mu_k \sim |q|^{1/2} \simeq (6\pi |\xi|)^{1/2}\Phi$ along
the physical  separatrix $A = \pi\Phi ^2 - 2|q|/3$.  Since the  
renormalized  $|\xi|$ may have very large values, this opens the way to 
exceptionally  efficient reheating - see Figs. 
(\ref{fig:nmc1},\ref{fig:nmc2}) - via resonant production of 
highly non-minimally  coupled fields with important consequences for  
GUT baryogenesis \cite{GPR97} and non-thermal  symmetry restoration.

%%%%%%%%%%%%%%%%%%%%%%%%%%%%%%%%%%%%%%%%%%%%%%%%%%
\bfg
\epsfxsize = 2.8in
\epsffile{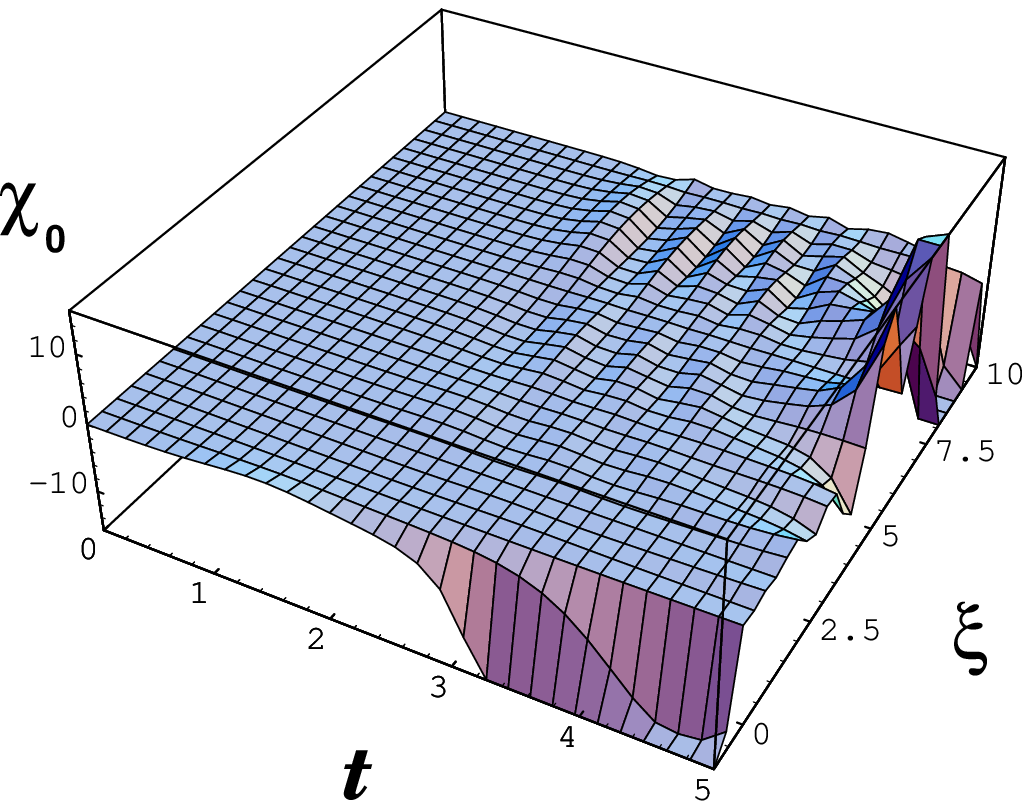}
\caption{
The evolution of the $k = 0$ mode ($m_{\nu}^2/m_{\phi}^2 \simeq 
1$),  as a function of time and the non-minimal coupling parameter $\xi$. 
For positive  $\xi$ the evolution is qualitatively that of the 
standard preheating with resonance bands. However, for negative 
$A$ (negative $\xi$) the solution changes 
qualitatively and there is a negative coupling instability. 
There are generically no stable bands and the Floquet index 
corresponding to $-|\xi|$ is much  larger,  scaling as 
$\mu_k \sim |\xi|^{1/2}$.
} 
\l{fig:nmc1} 
\efg

%%%%%%%%%%%%%%%%%%%%%%%%%%%%%%%%%%%%%%%%%%%%%%%%%%%%%%%%%%%%%%%%%%%%%%%%%%%%%%
\bfg
\epsfxsize = 2.8in
\epsffile{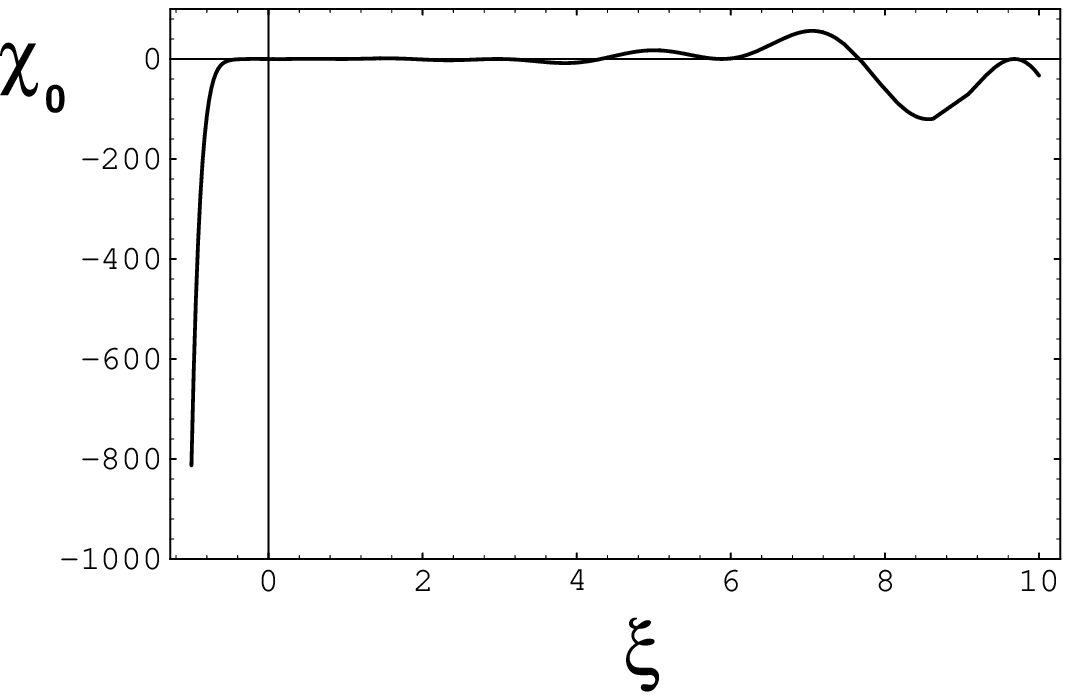}
\caption{A slice of the spectrum in fig. (1) at $t = 5$ as a 
function of the non-minimal coupling $\xi$. The qualitative differences 
between $\xi < 0 $ and $\xi > 0$ are clear.} 
\l{fig:nmc2} 
\efg
%%%%%%%%%%%%%%%%%%%%%%%%%%%%%%%%%%%%%%%%%%%%%%%%%%%%%%%%%%%%%%%%%%%%%%%%%%%

For example, let us consider $m_{\phi} \simeq 2\times 10^{13} GeV$ as 
required  to match CMB anisotropies $\Delta T/T \sim 10^{-5}$. Then GUT 
baryogenesis  with massive bosons $\chi$ with $m_{\chi} > 10^{14} GeV$ 
simply requires  $\xi < - (\pi \Phi^2)^{-1}$, with $\Phi$ in units of the 
Planck energy. Instead if one requires the production of GUT-scale gauge 
bosons  with masses $m_{gb} \sim 10^{16} GeV$ this is still possible if 
the associated non-minimal coupling is of order $\xi \sim -10^3$. Such 
coupling values have been considered in  e.g. \cite{SBB89}. The 
massive bosons with $m_{\chi} \sim 10^{14} GeV$ can be produced in the 
usual manner via parametric resonance  if $\xi > 0$, but 
this process is weaker (c.f. \cite{KLR96}). 

Since the coupling between $\phi$ and $\chi$ is purely gravitational, 
backreaction effects in the standard sense (see \cite{KLS97,devega}) cannot 
shut off the resonance. The inflaton continues to oscillate and 
produce non-minimally coupled  particles, receiving no corrections to
$m_{\phi, eff}^2$ from $\langle \delta \chi^2 \rangle$. 

To estimate  the maximum variance $\lan \delta 
\chi^2 \ran$ is therefore rather difficult. The standard method 
is to establish the time when the resonance is shut 
off by the growth of $A(k)$ which pushes the $k = 0$ mode out of the 
dominant  first resonance band. For this we must understand how $A(k)$ 
changes as the  $\chi$-field  gains energy and   
alters the Ricci curvature. If 
we assume that most of the energy goes into the $\chi_0$ mode, justified 
in the $\xi < 0$ case \footnote{In the case $\xi \gg 1$ one needs to use 
$\langle \delta \chi^2 \rangle$ instead.}, then the change to the Ricci 
curvature  is $\delta  R_{\chi} = 8 \pi(E - S)$, where \cite{marcelo}: 
\beq
E = G_{eff}\left[\frac{\dot{\chi}_0^2}{2} + 
\frac{m_{\chi}^2 \chi_0^2}{2} - 12 \xi \chi_0 \dot{\chi}_0 
\frac{\dot{a}}{a}\right]
\l{eq:nmce}
\eeq
is the $T^{00}$ component of the $\chi$ stress tensor,
\bea
S &=& \frac{3G_{eff}}{1 + 192\pi G_{eff} \xi^2 \chi_0^2} 
\left[\frac{\dot{\chi}_0^2}{2} - \frac{m_{\chi}^2 
\chi^2_0}{2} \right. \nonumber \\ 
&+&  \left. 4\xi \left( \frac{\dot{a}^2}{a^2} - \chi_0 
\dot{\chi}_0 \frac{\dot{a}}{a} -  m_{\chi}^2 \chi_0^2 \right) + 64\pi  
\xi^2 \chi_0^2 E \right] \,,
\l{eq:nmcs} 
\eea
is the spatial trace of the stress tensor $T^i{}_i$ corresponding 
to $3p$ in the perfect fluid case and 
$G_{eff} = (1 + 16\pi \xi \chi^2)^{-1} $
is the effective gravitational constant.  Since $\chi_0$ is  rapidly
growing, 
the major contribution of $\delta R_{\chi}$ will be to $A(k)$, causing a 
rapid vertical movement on the instability chart. Once $\delta A + A > 2|q| 
+ |q|^{1/2}$, the resonance is shut-off. If $\xi < 0$, most of the
decaying $\phi$ energy is
pumped into the small $k$ modes (see fig. \ref{fig:nmc2}). Subsequently we
expect the oscillations in $\chi_0$ to produce a secondary resonance due
to the self-interaction and non-linearity of Eq's
(\ref{eq:nmce},\ref{eq:nmcs}).  

The case of a $\lambda_{\chi} \chi^4/4$ self-interaction
provides another mechanism that may be dominant in ending the
resonance: namely $m_{\phi, eff}^2$, and hence $A(k)$, gains  corrections
proportional to $\lambda_{\chi} \lan \delta \chi^2 \ran$ 
which shuts off the resonance \cite{GPR97} leaving a peak variance of
order $\lan  \delta \chi^2 \ran \simeq m_{\phi}^2 (4|q| - 
m_{\chi}^2)/\lambda_{\chi}$ (assuming that $\xi < 0$). If 
$\xi > 0$ the variance is smaller by a factor $|q|^{1/2}$.

\section{The vector case}

Until now, reheating studies have been limited to minimally-coupled  scalar 
fields, fermions and gauge bosons \cite{BHP97}. In the case of vector fields
the minimum one can do to preserve gauge-invariance is to 
couple to a complex scalar field via 
the current since real scalar fields carry no quantum numbers.  
We consider here only  vacuum vector  resonances, however.

A massive spin--1 vector field 
in curved spacetime satisfies the equations:
\beq
(-\nabla_a \nabla^a + m_{\aa}^2)\aa^{b} + R^{b}{}_{a} \aa^a = 0
\l{eq:vec1}
\eeq
These equations are equivalent to the Maxwell-Proca equations for the vector
potential $\aa_a$ only after an appropriate gauge choice which removes one 
unphysical polarization. 
In our case we shall use the so-called tridimensional transversal 
constraint: 
\beq
\aa_0 = 0,~~~ \nabla^i \aa_i = 0
\l{eq:vec2}
\eeq
This set is equivalent to the Lorentz gauge,
although it doesn't conserve the covariant form of the latter.
Nonetheless, in either case, gauge-invariant quantities such as 
the radiation energy density, are unaffected.

In  a {\sc FLRW} background, the Ricci tensor is diagonal, which together
with the gauge choice (\ref{eq:vec2}) and expansion over eigenfunctions, 
ensures the decoupling of  the set of equations (\ref{eq:vec1}). We can
reduce the system to a  set of decoupled Mathieu equations. The Ricci
tensor is (see Eq. \ref{eq:ricci}): 
\beq
R^a{}_b = \kappa V(\phi) \delta^a{}_b - \kappa \dot{\phi}^2 \delta^a{}_0 
\delta^0{}_b\,, 
\l{eq:ricci1}
\eeq
which leads to the Mathieu  parameters for 
for the spatial components 
$(a^{3/2} \aa^i)$: 
\beq
A(k) \doteq \frac{k^2}{a^2 m_{\phi}^2} + \frac{m_{\aa}^2}{m_{\phi}^2} +  
2q~~~,~~ q \doteq \frac{\kappa \Phi^2}{8} 
\l{eq:spat}
\eeq
showing that vector fields are also parametrically amplified, albeit
weakly, during reheating as in the scalar case.

\section{The graviton case}

It has been shown using the electric 
and magnetic parts of the Weyl tensor \cite{bass97} that there exists
a formal analogy between the scalar field and graviton 
cases during resonant reheating.  Here we will show that the
correspondence  also holds in the Bardeen 
formalism. The gauge-invariant (at first order) transverse-traceless (TT)
metric perturbations $h_{ij}$  describe gravitational waves in the 
classical limit. 
In the Heisenberg picture one expands over eigenfunctions, $Y_{ab}$ of the 
{\em tensor} Laplace-Beltrami operator with scalar mode functions $h_k$,
which satisfy the equation of motion: 
\beq
\ddot{h}_k + \Theta \dot{h}_k + \left(\frac{k^2 + 2K}{a^2}\right) h_k = 0\,,
\l{eq:bard}
\eeq
or equivalently
\beq
(a^{3/2} h_k)\ddot{} + \left(\frac{k^2 + 2K}{a^2} + \ts{3\over4} 
p\right)(a^{3/2} h_k) = 0\,, 
\l{eq:bard2}
\eeq
where $p = \kappa(\dot{\phi}^2/2 - V)$ is the pressure. 
This gives a time-dependent Mathieu equation (c.f. Eq. \ref{eq:mcparam})  
with parameters: 
\beq 
A(k) \doteq \frac{k^2}{a^2 m_{\phi}^2}~~,~~q \doteq - \frac{3\kappa 
\Phi^2}{16}\,. 
\l{eq:bardmath}
\eeq 
In this case, a negative coupling instability is impossible and only for 
$\Phi \sim M_{pl}$ is there significant graviton production. Note, however, 
that if  temporal averaging is used, the average equation of state is 
that of  dust, $\overline{p} = 0$. Eq. (\ref{eq:bard2}) then 
predicts (falsely) that there is no resonant amplification of
gravitational waves since the value of $q$ corresponding to the
temporarily averaged evolution vanishes.

\section{Conclusions}

We have described a new -- geometric -- reheating channel after 
inflation, one  which 
occurs solely due to gravitational couplings. While this is not very 
strong in the  gravitational wave  and minimally coupled scalar field 
cases, it can be very powerful in the non-minimally coupled case, either 
due to broad-resonance ($\xi \gg 1$) or negative coupling 
($\xi < 0$) instabilities. Particularly in the latter case, it is 
possible to produce large numbers of bosons which are significantly  more 
massive than the inflaton, as required for GUT 
baryogenesis. It further gives rise to the possibility that the 
post-inflationary universe may be dominated by non-minimally coupled 
fields. These must be treated as imperfect 
fluids which would thus alter  both density perturbation and background 
spacetime evolution, which are known to be significantly different 
\cite{HM94} than in the simple perfect fluid case.
We have further presented a unified approach to resonant production of 
vector and tensor fields during reheating in analogy to the scalar case. 
  
Future work should examine in greater detail the possibility of 
negative coupling instability in the vector field, backreaction issues in 
the non-minimal case, and the situation in  potentials with 
self-interaction relevant to symmetry restoration. 

\section*{Acknowledgements}
We would like to thank Francesco Belgiorno, Marco Bruni, David Kaiser,
Marcelo Salgad, Dennis Sciama and Matt Visser for very useful comments and
discussions.

%%%%%%%%%%%%%%%%%%%%%%%%%%%%%%%%%%%%%%%%%%%%%%%%%%%%%%%%%%%%%%%%%%%%


\begin{references}


\bibitem{CL96} 
N. J. Cornish and J. J. Levin, 
Phys. Rev. D {\bf 53}, 3022 (1996)
 
\bibitem{hybrid} 
A. Linde, A. Riotto,  
%{\em Hybrid inflation in Supergravity},
Phys. Rev. D {\bf 56}, 1841 R (1997),  hep-ph/9703209;
%J. Garcia-Bellido, A. Linde, 
%{\em Open Hybrid Inflation},
%hep-ph/9701173 (1997) ;
%G. Dvali, L. M. Krauss, H. Liu, 
%{\em Hybrid inflation and particle  physics},  
%hep-ph/9707456 (1997);  
A.D. Linde, {\it Particle Physics and Inflationary Cosmology}
Harwood, Chur, Switzerland, (1990).


\bibitem{KLS97}  
L. Kofman, A. Linde and A. A. Starobinsky,
%{\em Towards the Theory of Reheating After Inflation},
Phys. Rev. D {\bf 56}, 3258 (1997), hep-ph/9704452 

\bibitem{pre1} 
L. Kofman, A. Linde and A. A. Starobinsky, 
Phys. Rev. Lett.  {\bf 73}, 3195 (1994);
Y. Shtanov, J. Traschen, and R. Brandenberger, 
Phys.Rev. D {\bf 51}, 5438 (1995).
H. Fujisaki, K. Kumekawa, M. Yamaguci and M. Yoshimura,
Phys. Rev. D {\bf 53}, 6805 (1996);


\bibitem{pre2} 
D. Kaiser, Phys. Rev. D {\bf 53}, 1776 (1996); 
T. Prokopec, T. G. Roos,
Phys.Rev. D {\bf 55}, 3768 (1997); 
S. Yu. Khlebnikov, I. I. Tkachev,
%{\em Resonant decay of Bose condensates},
hep-ph/9610477; 
S. Yu. Khlebnikov, I. I. Tkachev,
Phys. Rev. Lett. B {\bf 390}, 80 (1997).

\bibitem{devega} D. Boyanovsky, H. J. de Vega, R. Holman, J. F. J. Salgado, 
Phys.Rev.  D{\bf 54} 7570 (1996); 
D. Boyanovsky, H.J. de Vega, R. Holman, 
D.-S. Lee, and A. Singh, Phys. Rev. D {\bf 51}, 4419 (1995)

\bibitem{GKLS97} P. Greene,  L. Kofman, A. Linde, and A. A. Starobinsky,
%{\em Structure of Resonance in Preheating after Inflation},
hep-ph/9705347 (1997);


%\bibitem{GKLS97} 

\bibitem{BOS92} 
I.L. Buchbinder, S. D. Odintsov, I.L. Shapiro, 
{\em Effective Action in Quantum Gravity}.
IOP, Bath, 1992. 

\bibitem{RH97} S. A. Ramsey and B. L. Hu, Phys. Rev. D {\bf 56} 661 
(1997)  

\bibitem{BDH94} D. Boyanovsky, H. J. de Vega, R. Holman, 
Phys. Rev. D {\bf 49}, 2769 (1994)

\bibitem{PT} L. Parker and D. J. Toms, Phys. Rev. Lett. {\bf 
52}, 1269 (1984); {\em ibid} Phys. Rev. D {\bf 29}, 1584 (1984)

%\bibitem{devega} 
%D. Boyanovsky, M. D'Attanasio, H. J. de Vega, R. Holman, 
%D.S. Lee, and A. Singh, Phys. Rev. D {\bf 52}, 6805 (1995)

\bibitem{mac} 
N.W. McLachlan,
``{\em Theory and Application of Mathieu Functions}'',
Dover Publications, New York (1961).

\bibitem{GMM94} 
A. A. Grib, S. G. Mamayev, V. M. Mostepanenko,
``{\em Vacuum Quantum Effects in Strong Fields}'',
Friedmann Laboratory Publishing, St.Petersburg (1994).

%\bibitem{Y96} 
%M. Yoshimura,
%Prog. Theor. Phys. {\bf 94}, 873 (1995);

\bibitem{GPR97} 
B. R. Greene, T. Prokopec and T. G. Roos,
%{\em Inflaton Decay and Heavy Particle Production with Negative 
%Coupling}, 
hep-ph/9705357 (1997)

\bibitem{KS84} H. Kodama and M. Sasaki, Prog. Theo. Phys. Supp. {\bf 78}, 
1 (1984)

\bibitem{ellis71} G. F. R. Ellis, {\em Relativistic Cosmology},
in {\em Carg\'ese  Lectures in
Physics}, vol. VI, ed. E. Schatzmann (Gordon and Breach, 1973), p.1

\bibitem{KH96} H. Kodama, T. Hamazaki, Prog. Theor. Phys. {\bf 96}, 949 
(1996)

\bibitem{bass97} 
B. A. Bassett, Phys. Rev. D {\bf 56}, 3439 (1997); hep-ph/9704399
 
\bibitem{matacz} 
D. Koks, B. L. Hu, A. Matacz, A. Raval,
%{\em Thermal Particle Creation in Cosmological Spacetimes: A Stochastic
%Approach}, 
gr-qc/9704074 (1997)

\bibitem{SBB89} D.S. Salopek, J.R. Bond, J.M. Bardeen,
Phys. Rev. D {\bf 40}, 1753 (1989).

\bibitem{KLR96} 
E. W. Kolb, A. Linde, A. Riotto,
Phys. Rev. Lett. {\bf 77}, 4290 (1996). 

\bibitem{marcelo} M. Salgado, D. Sudarsky and H. Quevedo, Phys. Rev. 
D{\bf 53}, 6771 (1996)
 
%\bibitem{DBE97} 
%Dunsby, B. Bassett, G. Ellis, CQG {\bf 14}, 1215 (1997)  

\bibitem{BHP97} J. Baacke, K. Heitmann and C. Paetzold, Phys.Rev. D{\bf 
55}, 7815 (1997)

\bibitem{HM94} T. Hirai and K. Maeda, Astrophys. J.{\bf 431}, 6  (1994)

\end{references}
\end{document}